\documentstyle[12pt]{article}

\begin{document}

\author{M.\ Apostol  \\ 
Department of Theoretical Physics, \\Institute of Atomic Physics,
Magurele-Bucharest MG-6, \\POBox MG-35, Romania\\e-mail: apoma@theor1.ifa.ro}
\title{On a Thomas-Fermi model of ''hollow'' atom }
\date{J.\ Theor.\ Phys.{\bf 6\ }(1995)}
\maketitle

\begin{abstract}
A Thomas-Fermi model of a spherical shell of positive charge is
investigated, under various boundary conditions. The electron distribution
and the ionization charge are given particular attention.
\end{abstract}

\newpage
The advent of the fullerene molecule\cite{Cioslovsky} and the metallic
clusters\cite{Heer} have called the attention upon the stability of new
atomic micro-objects.\ One of them is described in this paper. As it is
known, the fullerene molecule $C_{60}$ consists of $60$ carbon atoms ($C$),
arranged (in pentagons and hexagons) on the surface of a sphere of radius $%
\sim 3.5\;\AA $. Certain clusters, made of (relatively) a small number of
atoms (as, for example, alkali atoms), have been identified in solid-state
matrices, particularly in the octahedral interstices of face-centered cubic
fullerites. They acquire regular geometrical shapes, like tetrahedrons, or
cubes, the latter being sometimes centered.\ We investigate herein the
stability of a Thomas-Fermi model suggested by these micro-objects.

As it is known, the Thomas-Fermi model starts with free electrons and
assumes that, due to the Pauli exclusion principle, there is a certain scale
length over which their number, and their effective interaction, vary
slowly. It is a quasi-classical theory, and its validity is ensured by the
number of electrons being much greater than unity. Of course, at infinitely
long distances the theory is not valid, nor at very short distances where
the Coulomb potential is singular. In the case of an atom with the atomic
number $Z$ we know that the Thomas-Fermi model is not valid for distances
shorter than $\sim 1/Z$. As in our case of the micro-objects presented above
we are interested in distances of the order of the size of large molecules,
or of the order of the solid-state distances, we may also neglect the
variations over atomic scale lengths, {\it i.e.} over the Bohr radius $%
a_H=\hbar ^2/me^2=0.53\;\AA $ (where $m$ is the electron mass, $e$ is the
electron charge and $\hbar $ is the Planck's constant).\ This allows one to
treat the atoms in the above micro-objects as being uniformly distributed
over a spherical shell of radius $R$. In addition, one may assume that the
centre of the sphere is occupied by a nucleus of positive charge $z_0$.

A positive charge $z$ uniformly distributed over the surface of a sphere of
radius $R$, and a central positive charge $z_0$, create an electric
potential 
\begin{equation}
\label{one}V_1(r)=z/R+z_0/r\;\;\;,\;\;\;r<R\;\;\;,
\end{equation}
and 
\begin{equation}
\label{two}V_2(r)=\left( z+z_0\right) /r\;\;\;,\;\;\;r>R\;\;\;.
\end{equation}
An electric field 
\begin{equation}
\label{three}E_s=\left( z/2+z_0\right) /R^2
\end{equation}
acts outwardly on the spherical surface, which tends to blow the sphere up;
and the electrostatic energy of the object is 
\begin{equation}
\label{four}{\cal E}_0=z\left( z/2+z_0\right) /R\;\;\;.
\end{equation}

The density of free electrons $n$ is related by the Fermi wavevector $k_F$
through $n=k_F^3/3\pi ^2$; the local energy of the electrons should be a
constant, for equilibrium, 
\begin{equation}
\label{five}\frac{\hbar ^2}{2m}k_F^2-e\varphi -eV=const\;\;\;, 
\end{equation}
where $\varphi $ is the electrostatic potential of the electron
distribution, and $V=V_{1,2}$ for $r<R$ and $r>R$, respectively. We shall
assume that the electron distribution extends to infinity, in which case $%
const=0$ in $\left( 5\right) $. We shall also use the atomic units $a_H$ and 
$e^2/a_H=27.2\;eV$, which together with $\hbar =1$ give $m=1$ and $e^2=1$.
Then, from $\left( 5\right) $, we have 
\begin{equation}
\label{six}n=\frac{2\sqrt{2}}{3\pi ^2}(\varphi +V)^{\frac 32} 
\end{equation}
and the Poisson equation reads 
\begin{equation}
\label{seven}\Delta \left( \varphi +V\right) =4\pi n=\frac{8\sqrt{2}}{3\pi }%
(\varphi +V)^{\frac 32} 
\end{equation}
for $r\neq R$; remark that $\Delta V=0$. Introducing the reduced variable $%
x=r/R$ and 
\begin{equation}
\label{eight}\varphi +V=\frac{9\pi ^2}{128R^4}\frac \chi x\;\;\;, 
\end{equation}
we get from $\left( 7\right) $ the Thomas-Fermi equation 
\begin{equation}
\label{nine}x^{\frac 12}\chi ^{^{\prime \prime }}=\chi ^{\frac 32}\;\;\;. 
\end{equation}
Since $n$ is a continuous function $\varphi $ and its two first derivatives
are continuous; therefore $\chi $ has a slope discontinuity (but itself is a
continuous function), exactly as the derivative of $V$. Defining $\chi
_1=\chi $ for $x<1$ and $\chi _2=\chi $ for $x>1$, we have therefore 
\begin{equation}
\label{ten}\chi _1(1)=\chi _2(1)\;\;\;. 
\end{equation}

The number $N(x)$ of electrons inside a sphere of radius $x$ is easily
obtained from 
\begin{equation}
\label{eleven}N(x)=4\pi R^3\int_0^xdx\;x^2n=\frac{9\pi ^2}{128R^3}%
\int_0^xdx\;\left( x\chi ^{^{\prime }}-\chi \right) ^{^{\prime }}\;\;\;, 
\end{equation}
whence 
\begin{equation}
\label{twelve}N_1(x)=\frac{9\pi ^2}{128R^3}\left[ x\chi _1^{^{\prime }}-\chi
_1+\chi _1(0)\right] \;\;\;,\;\;\;x<1\;\;\;, 
\end{equation}
and 
\begin{equation}
\label{thirteen}N_2(x)=\frac{9\pi ^2}{128R^3}\left\{ x\chi _2^{^{\prime
}}-\chi _2+\left[ \chi _1^{^{\prime }}(1)-\chi _2^{^{\prime }}(1)\right]
+\chi _1(0)\right\} \;\;\;,\;\;\;x>1\;\;\;. 
\end{equation}
We remark that $N(x)$ is continuous at $x=1$, and 
\begin{equation}
\label{fourteen}N(1)=\frac{9\pi ^2}{128R^3}\left[ \chi _1^{^{\prime
}}(1)-\chi _1(1)+\chi _1(0)\right] \;\;\;. 
\end{equation}
As the system extends to infinity we have to assume that $\chi _2(\infty )=0$
(together with its derivatives), so that the total number of electrons is
given by 
\begin{equation}
\label{fifteen}N=\frac{9\pi ^2}{128R^3}\left[ \chi _1^{^{\prime }}(1)-\chi
_2^{^{\prime }}(1)+\chi _1(0)\right] \;\;\;. 
\end{equation}

The electric field $E_1(x)$ at $x<1$, and the total charge $q_1(x)$ inside
the sphere of radius $x<1$, are obtained easily from the Gauss' law, 
\begin{equation}
\label{sixteen}4\pi r^2E_1=-4\pi Rx^2\frac \partial {\partial x}\left(
\varphi _1+V_1\right) =4\pi \frac{9\pi ^2}{128R^3}\left( \chi _1-x\chi
_1^{^{\prime }}\right) =4\pi q_1\;\;\;, 
\end{equation}
whence 
\begin{equation}
\label{seventeen}q_1=\frac{9\pi ^2}{128R^3}\left( \chi _1-x\chi _1^{^{\prime
}}\right) \;\;\;, 
\end{equation}
and 
\begin{equation}
\label{eighteen}E_1(x)=\frac{q_1(x)}{R^2x^2}\;\;\;. 
\end{equation}
Since $q_1=z_0-N_1$, we obtain from $\left( 12\right) $ and $\left(
17\right) $ 
\begin{equation}
\label{nineteen}z_0=\frac{9\pi ^2}{128R^3}\chi _1(0)\;\;\;. 
\end{equation}
Similarly, the electric field at $x>1$ is $E_2(x)=q_2(x)/R^2x^2$ and the
charge 
\begin{equation}
\label{twenty}q_2=\frac{9\pi ^2}{128R^3}\left( \chi _2-x\chi _2^{^{\prime
}}\right) \;\;\;, 
\end{equation}
whence, by using $\left( 13\right) $ and $\left( 19\right) $, we get 
\begin{equation}
\label{twentyone}z=\frac{9\pi ^2}{128R^3}\left( \chi _1^{^{\prime }}(1)-\chi
_2^{^{\prime }}(1)\right) \;\;\;. 
\end{equation}
We remark that $\left( 21\right) $ expresses exactly the jump in the slope
of $\chi $ and of $V$ at $x=1$, as we said above; and the total charge $%
q=q_2(\infty )=0,${\it i.e}${\cal .}$ the infinite system is neutral.

The electrons act on the shell with an electric field $E_{el}$ given by 
\begin{equation}
\label{twentytwo}
\begin{array}{c}
E_{el}=- 
\frac{\partial \varphi }{\partial r}\mid _{r=R}=-\frac{9\pi ^2}{128R^5}\frac
\partial {\partial x}\left( \frac{\chi _1}x\right) \mid _{x=1}+\frac{%
\partial V_1}{\partial r}\mid _{r=R}= \\ =-\frac{9\pi ^2}{128R^5}\left[ \chi
_1^{^{\prime }}(1)-\chi _1(1)\right] -\frac{z_0}{R^2}\;\;\;, 
\end{array}
\end{equation}
or, equivalently, 
\begin{equation}
\label{twentythree}E_{el}=-\frac{9\pi ^2}{128R^5}\left[ \chi _2^{^{\prime
}}(1)-\chi _2(1)\right] -\frac{z+z_0}{R^2}\;\;\;, 
\end{equation}
if we use $\left( 10\right) $ and $\left( 21\right) $. From $\left(
14\right) $ and $\left( 19\right) $ we find that 
\begin{equation}
\label{twentyfour}E_{el}=-N(1)/R^2\;\;\;, 
\end{equation}
as expected. In order to have the equilibrium of the shell this field must
compensate the field $E_s$ given by $\left( 3\right) $,{\it \ i.e. } 
\begin{equation}
\label{twentyfive}N(1)=z/2+z_0\;\;\;; 
\end{equation}
or, using $\left( 14\right) $, $\left( 19\right) $ and $\left( 21\right) $, 
\begin{equation}
\label{twentysix}2\chi _1(1)=\chi _1^{^{\prime }}(1)+\chi _2^{^{\prime
}}(1)\;\;\;. 
\end{equation}
Equations $\left( 19\right) $ and $\left( 21\right) $ may be viewed as
giving the parameters $z_0R^3$ and $zR^3$, respectively; therefore, we have
to solve the Thomas-Fermi equation $\left( 9\right) $ under the rather
natural conditions $\chi _1\left( 1\right) =\chi _2\left( 1\right) $, $\chi
_2\left( \infty \right) =0$ and $\left( 26\right) $. There is no such a
solution. We have always, in fact, $2\chi _1(1)>\chi _1^{^{\prime }}(1)+\chi
_2^{^{\prime }}(1)$, which means that $N(1)<z/2+z_0$,{\it \ i.e. }the
electrons inside the sphere are not numerous enough to ensure the
equilibrium; due to their fermionic nature they prefer to go outside the
sphere where their kinetic energy is lower.\ The infinite Thomas-Fermi
''hollow'' atom is too ''rarefied'' to be stable. Obviously, the only way to
attain the equilibrium of such a ''hollow'' atom is to embed it into a cage,
as, in fact, it is a more realistic case.

Suppose that we have a spherical cage of radius $R_0>R$, where the
''hollow'' atom is introduced. The Thomas-Fermi relationship $\left(
5\right) $ now reads 
\begin{equation}
\label{twentyseven}\frac{\hbar ^2}{2m}k_F^2-e\varphi -eV-U=-e\varphi
_0\;\;\;, 
\end{equation}
where $U$ is the potential well of the cage and $\varphi _0$ is the chemical
potential, which must equal the external potential for preventing the flux
of electrons from either going out or in the cage. Introducing the reduced
variable $x=r/R_0$ and defining 
\begin{equation}
\label{twentyeight}\varphi +V+U-\varphi _0=\frac{9\pi ^2}{128R_0^4}\frac
\chi x\;\;\; 
\end{equation}
(in atomic units) we arrive at the Thomas-Fermi equation $\left( 9\right) $
with th continuity condition $\left( 10\right) $ at $a=R/R_0$. The number of
electrons $\left( 12\right) $ and $\left( 13\right) $ is now given by 
\begin{equation}
\label{twentynine}N_1(x)=\frac{9\pi ^2}{128R_0^3}\left[ x\chi _1^{^{\prime
}}-\chi _1+\chi _1(0)\right] \;\;\;,\;\;\;x<a\;\;\;, 
\end{equation}
and 
\begin{equation}
\label{thirty}N_2(x)=\frac{9\pi ^2}{128R_0^3}\left\{ x\chi _2^{^{\prime
}}-\chi _2+a\left[ \chi _1^{^{\prime }}(1)-\chi _2^{^{\prime }}(1)\right]
+\chi _1(0)\right\} \;\;\;,\;\;\;x>a\;\;\;. 
\end{equation}
Similar relations $\left( 17\right) -\left( 20\right) $ hold now, with $R$
replaced by $R_0$; while the discontinuity condition $\left( 21\right) $
becomes now 
\begin{equation}
\label{thirtyone}z=\frac{9\pi ^2}{128R_0^3}a\left[ \chi _1^{^{\prime
}}(a)-\chi _2^{^{\prime }}(a)\right] \;\;\;. 
\end{equation}
A similar condition $\left( 25\right) $ for equilibrium is also obtained.
Summing up all these relationships we have to solve the Thomas-Fermi
equation $x^{\frac 12}\chi ^{^{\prime \prime }}=\chi ^{\frac 32}$ under the
following conditions: 
\begin{equation}
\label{thirtytwo}z_0=\frac{9\pi ^2}{128R_0^3}\chi _1(0)\;\;\;, 
\end{equation}
\begin{equation}
\label{thirtythree}\chi _1(a)=\chi _2(a)\;\;\;, 
\end{equation}
\begin{equation}
\label{thirtyfour}z=\frac{9\pi ^2}{128R_0^3}a\left[ \chi _1^{^{\prime
}}(a)-\chi _2^{^{\prime }}(a)\right] \;\;\;, 
\end{equation}
\begin{equation}
\label{thirtyfive}2\chi _1(a)=a\left[ \chi _1^{^{\prime }}(a)+\chi
_2^{^{\prime }}(a)\right] \;\;\;; 
\end{equation}
in which case the total charge in the cage is given by 
\begin{equation}
\label{thirtysix}q=\frac{9\pi ^2}{128R_0^3}\left[ \chi _2(1)-\chi
_2^{^{\prime }}(1)\right] \;\;\;. 
\end{equation}

First, we remark that, as we said above, leaving aside $\left( 32\right) $
and $\left( 34\right) $, as giving the parameters $z_0R_0^3$ and $zR_0^3$,
requiring $\chi _2(1)=0$ and letting $R_0$ go to infinity, we have the
infinite ''hollow'' atom discussed previously; and, then, it is easily to
see that the equilibrium condition $\left( 35\right) $ is not satisfied, as
it would require $\chi _1(a)=0$, {\it i.e.} a vanishing solution. Secondly,
we see that if we put $\chi _2(1)=0$ and $a=1$ we get again the previous
case of an infinite Thomas-Fermi atom, which we know that it is unstable; it
follows that even more unstable will be the ''hollow'' system with $\chi
_2(1)=0$ and $a<1$, {\it i.e} the ''positive ion''. But we remark that this
is only a particular cas of a positive ion.

In the remaining of this paper we shall discuss a few types of solutions for
the Thomas-Fermi equation $\left( 9\right) $ under the boundary conditions $%
\left( 32\right) -\left( 35\right) $, being especially interested in the
total charge $\left( 36\right) $.

For $z_0=0$, $R_0=2\;\AA $, $R=1.73\;\AA $ ($a=0.86$) and $z=44$, as for a
(tetrahedral) cluster of four sodium atoms, the function $\chi $ is plotted
in $Fig.1$ {\it vs} $x$; it corresponds to a total charge $q=+2.7$. For $%
z_0=0$, $R_0=3.2\;\AA $, $R=2.78\;\AA $ ($a=0.87$) and $z=88$, as for a
(cubic) cluster of eight sodium atoms, the function $\chi $ is shown in $%
Fig.2$; the total charge in this case is $q=-0.1$. For $z_0=11$, $%
R_0=3.15\;\AA $, $R=2.75\;\AA $ ($a=0.87$) and $z=88$, as for a centered
(cubic) cluster of nine sodium atoms, the function $\chi $ is given in $%
Fig.3 $, for a total charge $q=0.7$. We remark that, indeed, the function $%
\chi $ has a smoother variation with increasing the number of electrons,
except for a range $\sim a_H$ around the positions of atomic charges. In
addition, we remark that the charge $q$ is very sensitive to the input
parameters, for increasing both the number of electrons and the positive
charges.

We may estimate the energy of the system as follows. The density of kinetic
electron energy is 
\begin{equation}
\label{thirtyseven}\varepsilon _{kin}=\frac 2{\left( 2\pi \right) ^3}2\pi
\int_0^{k_F}dk\;\frac 12k^4=\frac{9(3\pi /32)^3}{10R_0^{10}}\frac{\chi ^{5/2}%
}{x^{5/2}}\;\;\;, 
\end{equation}
whence the total kinetic energy of electrons 
\begin{equation}
\label{thirtyeight}E_{kin}=\frac{3\left( 3\pi /8\right) ^4}{20R_0^7}%
\int_0^1dx\;\frac{\chi ^{5/2}}{x^{1/2}}\;\;\;. 
\end{equation}
The density of potential energy of the electrons is given by 
\begin{equation}
\label{thirtynine}\varepsilon _{pot}=-(\varphi +V+U)n=-\frac{3(3\pi /32)^3}{%
2R_0^{10}}\frac{\chi ^{5/2}}{x^{5/2}}-\frac{9\pi }{8^3R_0^6}\varphi _0\frac{%
\chi ^{3/2}}{x^{3/2}}\;\;\;, 
\end{equation}
whence their total energy 
\begin{equation}
\label{forty}{\cal E}_{el}=E_{kin}+E_{pot}=-\frac{\left( 3\pi /8\right) ^4}{%
10R_0^7}\int_0^1dx\;\frac{\chi ^{5/2}}{x^{1/2}}-N\varphi _0\;\;\;, 
\end{equation}
where $N=-(q-z-z_0)$ is the total number of electrons. The energy of the
shell is given by 
\begin{equation}
\label{fortyone}{\cal E}_s={\cal E}_0-(z+z_0)U\;\;\;. 
\end{equation}
On the other hand we may express the energy of interaction of the shell with
the electrons in two distinct ways

\begin{equation}
\label{fortytwo}{\cal E}_i=z\varphi \left( R\right) =-\int d{\bf r\;}%
nV\;\;\;, 
\end{equation}
where $V$ is the potential of the shell; from $\left( 42\right) $ we get 
\begin{equation}
\label{fortythree}U-\varphi _0=\frac{9\pi ^2}{128R_0^4}\chi _2^{^{\prime
}}\left( 1\right) \;\;\;, 
\end{equation}
which together with $\left( 40\right) $ and $\left( 41\right) $ allows one
to write the total energy as 
\begin{equation}
\label{fortyfour}{\cal E}={\cal E}_0+{\cal E}_1+{\cal E}_2-\left[ 2\left(
z+z_0\right) -q\right] \varphi _0\;\;\;, 
\end{equation}
where 
\begin{equation}
\label{fortyfive}{\cal E}_1=-\frac{\left( 3\pi /8\right) ^4}{10R_0^7}%
\int_0^1dx\;\frac{\chi ^{5/2}}{x^{1/2}} 
\end{equation}
and 
\begin{equation}
\label{fortysix}{\cal E}_2=-\frac{9\pi ^2}{128R_0^4}\left( z+z_0\right) \chi
_2^{^{\prime }}\left( 1\right) \;\;\;. 
\end{equation}
In the first case shown in $Fig.1$ ${\cal E}_1=-28$, ${\cal E}_2=-42$, ($%
\chi _2^{^{\prime }}\left( 1\right) =282$) while the self-energy $\left(
4\right) $ of the shell is ${\cal E}_0=297$; for stability, {\it i.e}.
negative total energy, one needs $\varphi _0\geq \left( 297-28-42\right)
/88\sim 2.6$. One can see that, indeed, the object is squeezed into the
atomic environment, $\varphi _0$ being a measure of the variation of the
atomic ''pseudo-potential'' felt by an outside electron on its attempts of
penetrating the electron cloud of the atomic surrounding; these
''pseudo-potentials'' are potential barriers which confine the clusters. A
huge ''pressure'' is exerted by the cluster on its surrounding, which
results in the deformation of the electronic clouds of the cage walls.
Similar values are obtained in the other cases, for example, $\varphi _0\geq
\left( 742-43-92\right) /176\sim 3.4$ for eight atoms, and $\varphi _0\geq
\left( 931-155-110\right) /200\sim 2.4$ for nine atoms.

It might be of interest the variation of $q$ with $z$. For example, we
define $Z=128R_0^3z/9\pi ^2$ and $Q=128R_0^3q/9\pi ^2$, and solve the
equation for various values of $a$. Such a dependence of $Q$ on $Z$ is shown
in $Fig.4$ for $z_0=0$ and $a=0.8$, and in $Fig.5$ for $z_0=0$ and $a=0.9$;
while in $Fig.6$ it is shown an almost neutral cluster for $z_0=0$ and $%
a=0.87$. Similar results can also be obtained for $z_0\neq 0$. In $Fig.7$
the variation of $q$ with $R_0$ is shown for $z_O=0$, $z=44$ and $%
R=1.73\;\AA $, while in $Fig.8$ a similar dependence is included for $z_O=11$%
, $z=88$ and $R=2.75\;\AA $.

In the limit of large number of atoms in the cluster the Thomas-Fermi theory
is valid.\ For a finite number $n$ of atoms disposed on a spherical surface
one may estimate the error in the total charge as follows. From the Poisson
equation we have that the charge $q$ is proportional to the radial $\delta
\varphi _r^{^{\prime }}$ and angular $\delta \varphi _a^{^{\prime }}$
variations of the potential derivatives in the following way: 
\begin{equation}
\label{fortyseven}q\sim \delta \varphi _r^{^{\prime }}\Delta S+2\delta
\varphi _a^{^{\prime }}\Delta S\;\;\;, 
\end{equation}
where $\Delta S$ is the element of area. Assuming the same variation per
unit length we find 
\begin{equation}
\label{fortyeight}\delta \varphi _a^{^{\prime }}=\delta \varphi _r^{^{\prime
}}\sqrt{\frac{4\pi }n}\;\;\;. 
\end{equation}
On the other hand, if one neglects the angular variations we have 
\begin{equation}
\label{fortynine}q_0\sim \left( \delta \varphi _r^{^{\prime }}\right)
_0\Delta S\;\;\;. 
\end{equation}
As these small variations are proportional to the small variations of the
distance we should also have 
\begin{equation}
\label{fifty}\left( \delta \varphi _r^{^{\prime }}\right) ^2+2\left( \delta
\varphi _a^{^{\prime }}\right) ^2=\left( 1+8\pi /n\right) \left( \delta
\varphi _r^{^{\prime }}\right) ^2=\left( \delta \varphi _r^{^{\prime
}}\right) _0^2\;\;\;, 
\end{equation}
whence 
\begin{equation}
\label{fiftyone}\Delta q/q\sim 1-\frac{\sqrt{1+8\pi /n}}{1+2\sqrt{4\pi /n}}%
\;\;\;. 
\end{equation}
In our cases of $n\sim 4,8,9$ this error in $\Delta q$ is about $40\%$.

\newpage\ 

Figure captions

$Fig.1$.

Function $\chi $ {\it vs} $x$ for $z_0=0$, $R_0=2\;\AA $, $R=1.73\;\AA $ ($%
a=0.86$) and $z=44$, and

a total charge $q=+2.7$.

$Fig.2$.

Function $\chi $ {\it vs} $x$ for $z_0=0$, $R_0=3.2\;\AA $, $R=2.78\;\AA $ ($%
a=0.87$) and $z=88$,

and a total charge $q=-0.1$.

$Fig.3$.

Function $\chi $ {\it vs} $x$ for $z_0=11$, $R_0=3.15\;\AA $, $R=2.75\;\AA $
($a=0.87$) and $z=88$,

and a total charge $q=0.7$.

$Fig.4$.

The reduced charge $Q$ {\it vs} $Z$ for $z_0=0$ and $a=0.8$.

$Fig.5$.

The reduced charge $Q$ {\it vs} $Z$ for $z_0=0$ and $a=0.9$.

$Fig.6$.

The reduced charge $Q$ {\it vs} $Z$ for $z_0=0$ and $a=0.87$ for an almost
neutral

cluster.

$Fig.7$.

The total charge $q$ {\it vs} $R_0$ for $z_0=0$, $z=44$, and $R=1.73\;\AA $.

$Fig.8$.

The total charge $q$ {\it vs} $R_0$ for $z_0=11$, $z=88$, and $R=2.75\;\AA $.

\end{document}